\documentclass[12pt]{article}

\usepackage{amsmath,amssymb,amsfonts}

\usepackage{times}
\usepackage{bm,mathrsfs,dsfont,color}
\usepackage{natbib}
\usepackage{graphicx}
\usepackage{subfigure}
\usepackage[font=small]{caption}

\usepackage[hyphens]{url} 
\usepackage[colorlinks=false, hidelinks]{hyperref}

\usepackage{authblk}

\allowdisplaybreaks

\usepackage{multirow}

\usepackage[plain,noend]{algorithm2e}

\makeatletter
\renewcommand{\algocf@captiontext}[2]{#1\algocf@typo. \AlCapFnt{}#2} 
\def\@algocf@capt@plain{top}
\renewcommand{\algocf@makecaption}[2]{%
  \addtolength{\hsize}{\algomargin}%
  \sbox\@tempboxa{\algocf@captiontext{#1}{#2}}%
  \ifdim\wd\@tempboxa >\hsize
    \hskip .5\algomargin%
    \parbox[t]{\hsize}{\algocf@captiontext{#1}{#2}}
  \else%
    \global\@minipagefalse%
    \hbox to\hsize{\box\@tempboxa}
  \fi%
  \addtolength{\hsize}{-\algomargin}%
}
\makeatother

\begin{document}

\makeatletter
\renewcommand\section{\@startsection {section}{1}{\z@}%
                                   {-3.5ex \@plus -1ex \@minus -.2ex}%
                                   {2.3ex \@plus.2ex}%
                                   {\centering\normalfont\large\scshape}}
                                   
\renewcommand\subsection{\@startsection {subsection}{1}{\z@}%
                                   {-3.5ex \@plus -1ex \@minus -.2ex}%
                                   {2.3ex \@plus.2ex}%
                                   {\centering \normalfont \scshape}}                                   
\makeatother

\title{Bayesian nonparametric analysis for the detection of spikes in noisy calcium imaging data}

\author[1]{
Laura D'Angelo\thanks{ laura.dangelo@unimib.it }}
\author[2]{  
Antonio Canale\thanks{ antonio.canale@unipd.it}}
\author[3]{
Zhaoxia Yu\thanks{ zhaoxia@ics.uci.edu}}
\author[3]{
Michele Guindani\thanks{ mguindan@uci.edu}
}
\affil[1]{Department of Economics, Management and Statistics; University of Milano-Bicocca}
\affil[2]{Department of Statistical Sciences; University of Padova}
\affil[3]{Department of Statistics; University of California, Irvine}

\date{}

\maketitle

\abstract{
Recent advancements in miniaturized fluorescence microscopy have made it possible to investigate neuronal responses to external stimuli in awake behaving animals through the analysis of intra-cellular calcium signals. An on-going challenge is deconvolving the temporal signals to extract the spike trains from the noisy calcium signals' time-series. In this manuscript, we propose a nested Bayesian finite mixture specification that allows the estimation of spiking activity and, simultaneously, reconstructing the distributions of the calcium transient spikes' amplitudes under different experimental conditions. The proposed model leverages two nested layers of random discrete mixture priors to borrow information between experiments and discover similarities in the distributional patterns of neuronal responses to different stimuli. Furthermore, the spikes' intensity values are also clustered within and between experimental conditions to determine the existence of common (recurring) response amplitudes. Simulation studies and the analysis of a data set from the Allen Brain Observatory show the effectiveness of the method in clustering and detecting neuronal activities. }

{\center \textbf{Keywords: }}
Dirichlet process; Mixtures of Finite Mixtures; Model-based clustering; Nested Dirichlet process; Spike and slab. \vfill

\section{Introduction}
\label{s:intro}

In recent years, calcium imaging has become a popular technique to measure the neuronal activity in awake, freely moving and behaving animals over time.
Due to the development of miniaturized and flexible microendoscopes for fluorescence microscopy, this technique has enabled the study of how individual neurons and neuronal networks encode external stimuli and cognitive processes \citep{Nakajima2020, li2015motor}.
Calcium ions generate intracellular signals that determine a large variety of functions in all neurons \citep{Grienberger2012}.
The mechanism at the basis of calcium imaging is a physiological process of the cells: when a neuron fires, calcium floods the cell and produces a transient spike in its concentration. By using genetically encoded calcium indicators, which are fluorescent molecules that react when binding to the calcium ions, it is possible to optically measure the level of calcium ions  by analyzing the observed fluorescence trace. 
The outcome of this technique is a movie of time-varying fluorescence intensities, from which the spike trains of the observable neurons are often extracted through a complex pre-processing phase. In general, this phase is meant to deal with two issues: identifying the spatial location of each neuron in the optical field, and deconvolving the temporal signals to extract their spike trains. Several methods are employed to extract the fluorescence traces, e.g. by using independent component analysis~\citep{Dombeck2010, Mukamel2009}, non-negative deconvolution~\citep{vogelstein2010} and non-negative matrix factorization~\citep{Maruyama2014}.

The resulting processed data consist of a fluorescent calcium trace for each observable neuron in the targeted area (see Figure \ref{fig:y} for an example). 
Although the observed fluorescence trace can be analyzed directly~\citep{shen2021},  it is a proxy of the underlying cellular activity and the information relevant to many studies often requires 
the precise spike times and the intracellular calcium concentration of the observable neurons when the animal is subjected to external stimuli~\citep{vogelstein2009}.
Extracting the neuronal activity from these series is not trivial: the calcium imaging technology has several limitations, including the presence of measurement noise, the nonlinearity between fluorescence transient and calcium concentration and the slow decay of the fluorescence trace compared to the underlying neuronal activity~\citep{friedrich2017,dana2019, Rose2014}. Moreover, the large-scale of the time series introduces additional complexity to the analysis. Therefore, a precise estimation of the spike times and amplitudes is a fundamental step toward the understanding of the neurons' behavior.

As a motivating application, here we consider a publicly available data set from the Allen Brain Observatory~\citep{allen, vries2020} of calcium imaging data obtained through two-photons microscopy in behaving mice. This study is an extended \textit{in vivo} survey of physiological activity in the mouse visual cortex in response to a range of visual stimuli~\citep{allen_stimulus}. 
Each mouse is placed in front of a screen where different types of visual stimuli are shown, while the mouse's neuronal activity is recorded. 
The stimuli vary from simple synthetic images such as locally sparse noise or static gratings, to complex natural scenes and movies.
The goal of the study is to investigate how neurons at different depths in the visual areas respond to stimuli of different complexity. Specifically, each neuron in the visual cortex can be characterized by their \textit{receptive field}, i.e. the features of the visual stimulus that trigger the signaling of that neuron.
Hence, it is of critical interest to devise methods that allow inferring how the neuronal response varies under the different types of visual stimuli. We expect that the neuronal activity will vary across all the experimental settings, and that some variations in its intensity will be observed based on the stimulus. 

Several approaches have been proposed to accurately and efficiently estimate the neuronal activity in calcium imaging data from single neurons. For example,~\citet{friedrich2016} and~\citet{friedrich2017} have proposed an on-line algorithm based on a lasso penalty to enforce sparsity of signal detection. \citet{jewell2018} and~\citet{jewell2019} have proposed using an $L_0$ penalty in lieu of the $L_1$ penalization and an efficient algorithm to identify the presence or absence of spikes. In a Bayesian framework, \citet{pnevmatikakis2013} have proposed to conduct inference on spike trains by estimating posterior probabilities of a latent binary indicator of spike presence at each time point. 
However, the model in \citet{pnevmatikakis2013} does not explicitly assume sparsity of the spikes. Moreover, it is expected the rate and the distribution of spikes to be stimulus-dependent \citep{Brenner2002PhysRevE}, but none of the previous approaches accounts for the heterogeneity of spikes' behaviors as a function of the stimulus. 
As Figure~\ref{fig:y} clearly shows for the Allen Brain Observatory data, the spikes' intensities vary greatly according to the type of stimulus. See also the discussion in \citet{Shibue2020} where they employ a marked point processes for the deconvolution of calcium imaging data.

In this manuscript, we introduce a coherent nested Bayesian finite mixture model that allows the estimation of the the spiking activity of each neuron -- which could be seen as a first step for the analysis of larger brain activity combining multiple neurons in a region. In addition, our model \textit{simultaneously} allows us to  reconstruct the distributions of spikes under various experimental conditions; for example, in response to different types of visual stimuli in the Allen Brain Observatory data set. More specifically, our modeling framework estimates and clusters the distributions of the calcium transient spikes' amplitudes via a nested formulation of 
mixture of finite mixtures \citep{miller2018, argiento2019} and, in particular, exploiting the generalized mixture of finite mixtures (gMFM) prior recently proposed by \citet{fruhwirthschnatter2020}. 
The proposed model further adopts the use of a common atom specification as in \citet{denti2020} for estimating the distribution of the spikes' amplitudes under each experimental condition. The proposed common atom gMFM has several advantages with respect to typical Bayesian nonparametric models for nested data. With respect to models based on Dirichlet process priors, the gMFM provides increased flexibility to estimate partitions characterized either by many, well-balanced, clusters or by a small set of large clusters. The common atom model allows us to obtain  nested inference on densities without incurring in the degeneracy issues pointed out by \citet{camerlenghi2019} for the widely used nested Dirichlet process of \citet{Rodriguez2008}. At the same time, the common atom formulation still leverages two nested layers of random discrete mixture priors to borrow information between experiments and to identify similarities in the distributional patterns of the neuronal responses to different stimuli. In addition, differently than in the nested Dirichlet process, the common atom model  also allows us to cluster the inferred spikes' intensity values both within and between experimental conditions, so to infer common (recurring) response amplitudes. Finally, we allow our model to enforce sparsity of neuron firing over time by assuming a spike-and-slab prior specification on the marginal distribution of the amplitudes.

\section{Bayesian mixture model for calcium imaging data}

\subsection{Model and prior specification}
\label{s:model}

The observed fluorescence is often considered as a noisy realization of the underlying true calcium concentration. To model a neuron's activity, we adopt a popular model in the neuroscience literature, where the decay in fluorescence is modeled through an autoregressive process and the spikes are modeled as jumps in correspondence to the neuron's firing events~\citep{vogelstein2010}.
Denoting with $y_t$ the observed fluorescence trace of a neuron and with $c_t$ the underlying calcium concentration, for $t=1,\dots,T$, one can assume 
\begin{eqnarray}
&y_t = b + c_t + \epsilon_t,\quad \epsilon_t \sim \text{N}(0,\sigma^2), \nonumber \\
&c_t = \gamma c_{t-1} + A_t + w_t, \quad w_t \sim \text{N}(0, \tau^2),
\label{eq:armodel}
\end{eqnarray}
where $b$ models the baseline level of the observed trace and $\epsilon_t$ is a measurement error. In the absence of neuronal activity, the true calcium concentration $c_t$ is considered to be centered around zero. The parameter $A_t$ captures the neuronal activity: in the absence of a spike ($A_t = 0$), the calcium level follows a AR(1) process controlled by the parameter $\gamma$
; when a spike occurs, the concentration increases instantaneously with the spike amplitude $A_t > 0$.

We are interested in characterizing the neuronal activity under different experimental conditions. For each time point $t=1,\dots,T$, let $g_t$ be a discrete categorical variable, taking values in $\{1,\dots, J\}$, where $J$ is the number of distinct experimental settings, so that $g_t=j$ indicates that the neuronal activity at time $t$ is observed under condition $j$. The experimental conditions are often designed to capture variations in neuronal activity with respect to a baseline process, which may represent a ``typical" brain process. For example, in the Allen Observatory data, the interest is to investigate visually-evoked functional responses of neurons in the mouse's visual cortex. Therefore,  some  neurons associated with visual decoding should be expected to activate in all conditions. It is then of interest to study not only \textit{if} but also \textit{how} the neurons differentially respond to the presentation of a variety of visual stimuli. 

In this paper, we propose a hierarchical Bayesian approach to investigate similarities and differences in the distribution of spikes over time and conditions. In order to borrow information across different experimental conditions, one option is to fit a parametric hierarchical random effect model, and obtain a post-MCMC clustering of the estimated spikes $A_t$ by grouping together those spikes with similar magnitudes. This approach has several limitations: on the one hand, the distribution of the random effects is constrained into a specific parametric form; on the other hand, the clustering of, say, the posterior mean estimates of the parameters $A_t$'s does not allow the model to fully describe stimulus-specific distributional differences and to take into account the posterior uncertainty in the spikes.

In order to allow flexible modeling of distributions and to describe the heterogeneity of distributional features, we assume a nested Bayesian finite mixture specification. More specifically, we rewrite (\ref{eq:armodel}) as
\[
y_t\mid b, \gamma, c_{t-1}, A_t, \sigma^2, \tau^2 \sim \text{N}(b+\gamma\:c_{t-1} + A_t, \sigma^2 + \tau^2 )
\]
and we assume that the spikes $A_t$ are from stimulus-specific distributions, i.e. 
$ (A_t \mid g_t = j, \, G_j) \sim G_j$, $ j=1, \ldots, J$,
to account for the observed variety of neuronal activity under different experiment settings. We further allow  clustering the distributions across conditions, in order to capture similar patterns of neuronal activity. Indeed, one may typically expect $K<J$ distributional clusters. For example, a neuron may respond to general visual stimulation and not specifically to the type of stimulus considered. More specifically, we assume the following generalized mixture of finite mixtures structure:
\begin{equation}
G_1,\dots,\,G_J \mid Q \sim Q, \qquad Q = \sum_{k=1}^{K} \pi_k \delta_{G^*_k}
\label{eq:first_layer}
\end{equation}
where \(\pi_1,\dots,\pi_K \mid K \sim \text{Dirichlet}_K\left(\alpha/K, \ldots \alpha/K\right)\), $\alpha>0$, and $G_1^*, \ldots, G_K^*$ are a set of cluster-defining distributions, obtained as realizations of an underlying random probability measure, specified further below. Equation \eqref{eq:first_layer} implies that the $G_j$'s, $j=1,\ldots, J$ have a positive probability of clustering together, thereby giving rise to \textit{distributional clusters}. In practice, the number of mixture components, $K$, is typically larger than the number of clusters, $K_+$, and some of the atoms $G_k^*$ are not assigned to any of the $G_j$'s (empty components). The prior on the number of mixture components $K$ is a translated beta-negative-binomial distribution as in \citet{fruhwirthschnatter2020}. 
Including a prior $p(K)$ leads to both $K_+$ and $K$ being random a priori. 
Finally, the distributional atoms $G_k^*$, $k=1, \ldots, K$ are also obtained as a realization from an underlying generalized mixture of finite mixtures, 
\begin{equation}
G^*_k = \sum_{l=1}^{L} \omega_{l,k} \delta_{A^*_l}
\label{eq:second_layer}
\end{equation}
with $\omega_{1,k},\dots,\omega_{L,k} \mid L\sim \text{Dirichlet}_L\left(\beta/L\right)$, for some positive real number $\beta>0$. The set of atoms $A_l^*$ is common across all distributions $G_1^*, \ldots, G_K^*$ and they are obtained as i.i.d. draws from a centering measure, \(A^*_l \sim G_0(A^*_l)\). Therefore, equation \eqref{eq:second_layer} defines a clustering of the inferred spike intensities both within a given condition (i.e. for fixed $G_k^*$) and across conditions (i.e. across the $G_k^*$'s; hence, across the $G_j$'s). In the following, we adopt common terminology in the literature on nested Bayesian non-parametric priors and indicate the clustering induced on the $A_t$ through the proposed two-layers prior as \textit{observational clustering}. The nested gMFM formulation requires the specification of a prior on the number of components that specify the lower-level distributional atoms $G_k^*$, \(L\sim p(L)\). Once again, some of the components may be empty. 

We enforce sparsity in the detection of the spikes by modeling the base measure $G_0$ for the parameters $A^*_{l}$ with a spike-and-slab specification \citep{mitchell(88)}, which is a convex mixture between a Dirac mass at zero -- representing the absence of neuronal response -- and a diffuse density on the positive real numbers -- representing the intensity of the neuronal response. More specifically, we assume
\begin{equation}
G_0 = (1-p) \, \delta_0 + p\, \text{Ga}\,(h_{A1},h_{A2}),
\label{eq:G0}
\end{equation}
where the slab is a gamma distribution, $\text{Ga}(a,b)$ with mean $a/b$ and variance $a/b^2$. The choice of a gamma distribution in (\ref{eq:G0}) is particularly relevant for sparsity-inducing purposes, as the gamma density belongs to the set of moment non-local prior densities, as defined by~\citet{johnson2010}. Therefore, a negligible probability density is assigned to values in a neighborhood of zero, thus inducing a clear separation between the baseline neuronal activity and the neuronal responses. In particular, the higher the shape parameter $h_{A1}$, the larger is the separation. We assume a $ \text{Beta}(h_{1p}, h_{2p})$ prior for the proportion of spikes $p$ with $h_{1p} $ much smaller than $h_{2p}$ in order to favor sparsity of detections.
A spike-and-Gamma model has also been used for the analysis of calcium imaging data  by \citet{wei2019}, although in a two-stage set-up for modeling the distribution of the (already deconvolved) estimated spikes. 

The proposed formulation can be seen as a special case of \emph{inner} spike-and-slab nonparametric priors, following a terminology introduced by \citet{canale2017,spikeandslab2}. 
In the following, we will refer to the proposed specification as a finite common atom model (fCAM).

The Bayesian model elicitation is completed by assuming conjugate priors for the underlying calcium level concentration parameters, i.e. the baseline calcium level $b$, and the variances $\sigma^2$ and $\tau^2$.  Specifically, the following conjugate prior distributions are assumed:
\begin{eqnarray*}
	&c_0 \sim \text{N}(0,C_0), \quad b\sim \text{N}(b_0,B_0) \\ 
	&1/\sigma^2 \sim \text{Ga}(h_{1\sigma},h_{2\sigma}),\quad 
	1/\tau^2 \sim \text{Ga}(h_{1\tau},h_{2\tau}),
\end{eqnarray*}
Finally, under the assumption that the process is stationary with positive correlation between the calcium level at consecutive times, we constrain $\gamma \in (0,1)$ and let $\gamma\sim \text{Beta}(h_{1\gamma}, h_{2\gamma})$, a priori.

Consistently with our aim of modelling the activity of a single neuron, each unknown parameter is neuron-specific. However,  the  hyper-parameters of the priors above can be specified at population level, especially in case of independent analyses for multiple neurons.

\subsection{Posterior inference}
\label{s:posterior_inference}

For computational purposes, it is often convenient to rewrite the likelihood for an observation $y_t$ under condition $g_t=j$ by introducing two latent cluster allocation variables, $S_j = S_{g_t}$ and $M_t$, indicating the distributional cluster for the group $j$ and the observational cluster for $y_t$, respectively. 

Given $K$ and $\{\pi_k\}_{k=1}^K$, the distributional allocation variable $S_j \sim \text{Multin}_K(\bm{\pi})$, where $\text{Multin}_K$ denotes a multinomial distribution with $K$ categories and event probabilities $\bm{\pi}$. Similarly, given $L$ and $\{\omega_l\}_{l=1}^L$, the observational allocation variable $M_t \sim \text{Multin}_L(\bm{\omega})$. For notation simplicity, we have indicated the parameter vectors using bold letters in the equations above. 
Therefore, conditionally on the other model parameters, the joint distribution of the observed data and the latent cluster allocations can be written as
\[
f(\bm{y},\bm{M},\bm{S}\mid \bm{\pi}, \bm{\omega}, \bm{A}^*) = \prod_{j=1}^J \pi_{S_j} \prod_{t:g_t = j} \omega_{M_t,S_j}\: p(y_t\mid A^*_{M_t}),
\]
which facilitates posterior inference. 

More specifically, posterior inference for the proposed fCAM can be carried out quite straightforwardly by means of Markov chain Monte Carlo (MCMC) techniques. The sampling of the latent calcium level $c_t$ uses an iterative approach based on the Kalman filter and on a forward filtering backward sampling algorithm \citep{prado2010}.
Full conditional posteriors for $b$, $p$, $\sigma^2$ and $\tau^2$ are available in closed form thus leading to straightforward Gibbs sampling steps. For the autoregressive parameter $\gamma$, we use a Metropolis-Hastings within the Gibbs step. 
The sampling of $A_t$ exploits a combination of the nested slice sampler of~\citet{denti2020} and of the telescoping sampler of~\citet{fruhwirthschnatter2020}. A detailed description of the latter step is reported in the the Appendix. Here, we just present a schematic description of the MCMC steps:
\begin{enumerate}
	\item[1)] Sample the calcium level $c_t$, for $t=0,\dots,T$, using a forward filtering backward sampling:
	\begin{itemize}
		\item[a)] Run Kalman filter: set $a_0 = m_0 = 0$, $R_0 = C_0 = \text{var}(c_0)$. For $t = 1,\dots,T$ let 
		\[
		a_t = \gamma \, m_{t-1} + A_t
		\]
		\[
		R_t = \gamma^2 \, C_{t-1} + \tau^2.
		\]
		Compute the filtering distribution's parameters, $m_t$ and $C_t$, for $t = 1,\dots,T$, where
		\[
		m_t = a_t + R_t\, (R_t + \sigma^2)^{-1} \, (y_t - b - a_t)
		\]
		\[
		C_t = R_t -  R_t^2 \, (R_t + \sigma^2)^{-1}.
		\]
		\item[b)] Draw $c_T \sim \text{N}(m_T, C_T)$;
		\item[c)] For $t = T-1, \dots, 0$, draw $c_t \sim \text{N}(h_t, H_t)$, with 
		\[
		h_t = m_t + \gamma \, C_t \, R_{t+1}^{-1} (c_{t+1} - a_{t+1})
		\]
		\[
		H_t = C_t - \gamma^2 \, C_t^2 \, R_{t+1}^{-1}.
		\]
	\end{itemize}
	\item[2)] Sample a new value for the baseline level $b$: 
	\[
	b  \sim \text{N}\left( \frac{b_0}{B_0} + \frac{1}{\sigma^2} \sum_{t=1}^T(y_t - c_t), \sqrt{\frac{1}{B_0} + \frac{T}{\sigma^2}} \right).
	\]
	\item[3)] Sample the variance on the output equation $\sigma^2$ and the variance on the state equation $\tau^2$:
	\[
	1/\sigma^2 \sim \text{Ga}\left( h_{1\sigma} + \frac{T}{2},\, h_{2\sigma} + \frac{1}{2} \sum_{t=1}^T (y_t - c_t - b)^2 \right)
	\]
	\[
	1/\tau^2 \sim \text{Ga}\left( h_{1\tau} + \frac{T}{2},\, h_{2\tau} + \frac{1}{2} \sum_{t=1}^T (c_t - \gamma \, c_{t-1} - A_t)^2 \right).
	\]
	\item[4)] Update the autoregressive parameter $\gamma$ using a Metropolis-Hastings step.
	\item[5)] Update the parameter $p$ of the spike-and-slab base measure from
	\[
	p \sim \text{Beta}(h_{1p} + T - n_0,\, h_{2p} + n_0), 
	\]
	where $n_0$ is the number of $y_t$ assigned to the the spike component.
	\item[6)] Update the cluster allocations variables $\bm{S}$ and $\bm{M}$, the number of mixture components $K$ and $L$, and the cluster parameters $\bm{A^*}$ using the nested telescoping sampling for the finite common atom model reported in the Appendix.
	
\end{enumerate}

\section{Simulation study}
\label{s:simulations}   

The performances of the proposed method are assessed through a simulation study. The purpose of this section is twofold, namely to assess both the ability to correctly identify the spike times, and the accuracy of the inferred clustering structure.  


We simulated synthetic data exhibiting a baseline level and a number of spikes representing the effect of the response of a neuron to a stimulus, thus mimicking the characteristics of real series of calcium imaging following the structure of model (\ref{eq:armodel}).
Specifically, we first divided the time frame into $J$ hypothetical experimental conditions of equal length, with $J$ varying in the different scenarios described below. Consistent with our motivating assumption that the neuronal response depends on the type of stimulus, each experimental condition is assumed to belong to one of the $K$ distributional clusters. Then, for each experimental condition, we generated the neuronal activity: first, we generated the presence or absence of a neuron response uniformly in time, where the spike probability can vary across groups. 
Then, conditionally on the obtained activations, we generated some additional spikes in a short subsequent interval, so that it is very likely to observe close or even successive spikes.
In this way, the data mimic a real calcium imaging time series. Moreover, we are able to conduct a careful assessment of the ability of the model to distinguish the presence of a single high spike versus the convolution of several spikes in consecutive times.
Finally, the values $A_t$, conditionally on their distributional cluster, are generated from one of the finite sets of spike amplitudes described below.

\begin{figure}
	\centerline{
		\includegraphics[width = .32\linewidth]{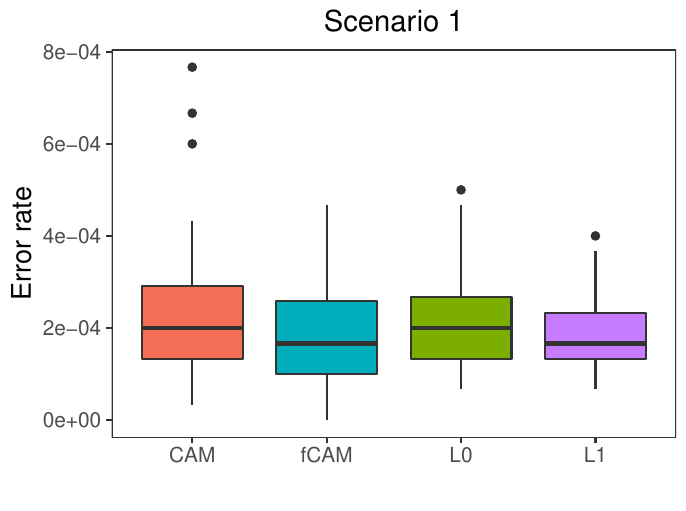}
		\includegraphics[width = .32\linewidth]{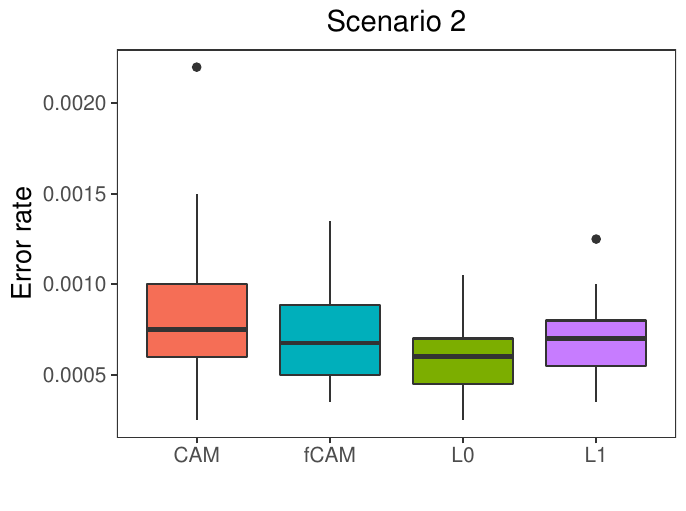}
		\includegraphics[width = .32\linewidth]{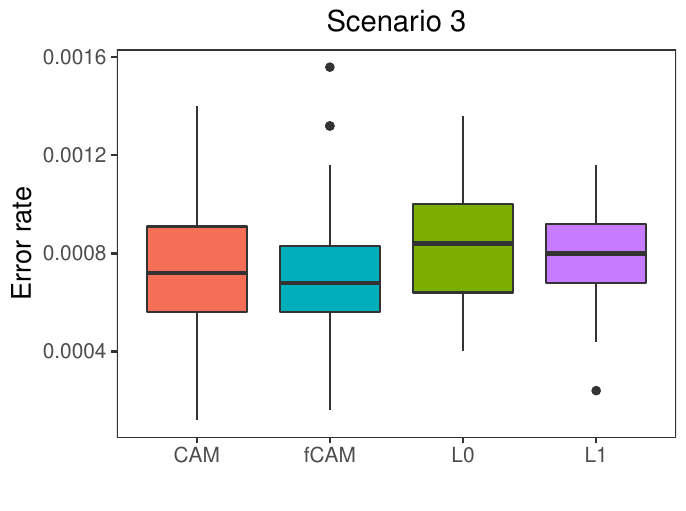}
	}
	\caption{Distribution of the misclassification error rate in the simulation study for the four considered methods: CAM, fCAM, and the methods of~\citet{jewell2019} ``L0'' and~\citet{friedrich2017} ``L1''. This figure appears in color in the electronic version of this article, and any mention of color refers to that version.}
	\label{fig:boxplots_rates}
\end{figure}

We simulated 50 independent data sets for each of the three scenarios described henceforth. In Scenario 1 we assumed $J=6$ experimental conditions, generated from $K=4$ distributional clusters. 
The spike amplitudes in the distributional clusters are (0.35, 0.89, 1.15, 1.80, 2.20), (0.65, 0.89, 1.40, 1.80), (0.35, 0.65, 1.15), and (0.35, 0.89, 1.60).
Scenario 2 assumes $J=4$ experimental conditions and $K=3$ distributional clusters with spike amplitudes equal to (0.3, 0.5, 0.7, 0.9, 1.1, 1.5), (0.3, 0.9, 1.5, 1.8), and (0.5, 0.9, 1.5). 
Finally, Scenario 3 sets $J=5$ and $K=3$ with the spike amplitudes in the distributional clusters being (0.3, 0.5, 0.7, 0.9, 1.1), (0.3, 0.9, 1.1, 1.3), and (0.7, 0.9, 1.3). 
While in Scenario 1 the amplitudes of the spikes are quite large, spaced apart, and with the corresponding distributional clusters well distinct, in Scenario 3 the spike amplitudes are more homogeneous and more clustered in time. Scenario 2 represents an in-between situation. Hence, from the first to the last scenario, we are assuming an increasing degree of complexity. The R script generating these synthetic datasets is described in the Supporting Information.

The results attained by the proposed fCAM are compared to those obtained exploiting the common atom model (CAM) of~\citet{denti2020} -- which provides a benchmark for the clustering of the spikes and the stimulus-specific distributions -- and to those obtained with the $L_0$ penalization method of~\citet{jewell2019} and the $L_1$ penalization method of~\citet{friedrich2017}, which provide a benchmark for the task of spikes' detection. For the latter two methods we have assumed complete knowledge of the autoregressive constant controlling the rate of the calcium decay, since we found that the results were quite sensitive to this estimate. 

To assess the sensitivity of the proposed fCAM to the prior specification, we repeated the numerical experiment for different values of the hyperparameters $h_{A1}$ and $h_{A2}$ in \eqref{eq:G0}. In particular, the shape parameter $h_{A1}$ was supposed to play a key role in the detection of spikes. Keeping fixed the ratio $h_{A1}/h_{A2}$, the parameters were set equal to 3, 4, 6, and 8: a small value implies, \emph{a priori}, less separation between zero and the distribution of the positive spikes while a large value corresponds to the opposite effect.

Focusing on the classification of each time point as a spike or not, Figure~\ref{fig:boxplots_rates} summarizes the misclassification rate for all competing methods under the three scenarios. The results of the $50$ replications are summarized using boxplots. For our fCAM, we report only the results obtained with $h_{A1} = h_{A2} = 8$ as those obtained for the other choices are essentially equivalent.  The rates are small in absolute value and broadly comparable across the different methods, thus confirming that all the competing models are effective in detecting the spikes.  

\begin{figure}
	\centerline{\includegraphics[width = .6\linewidth]{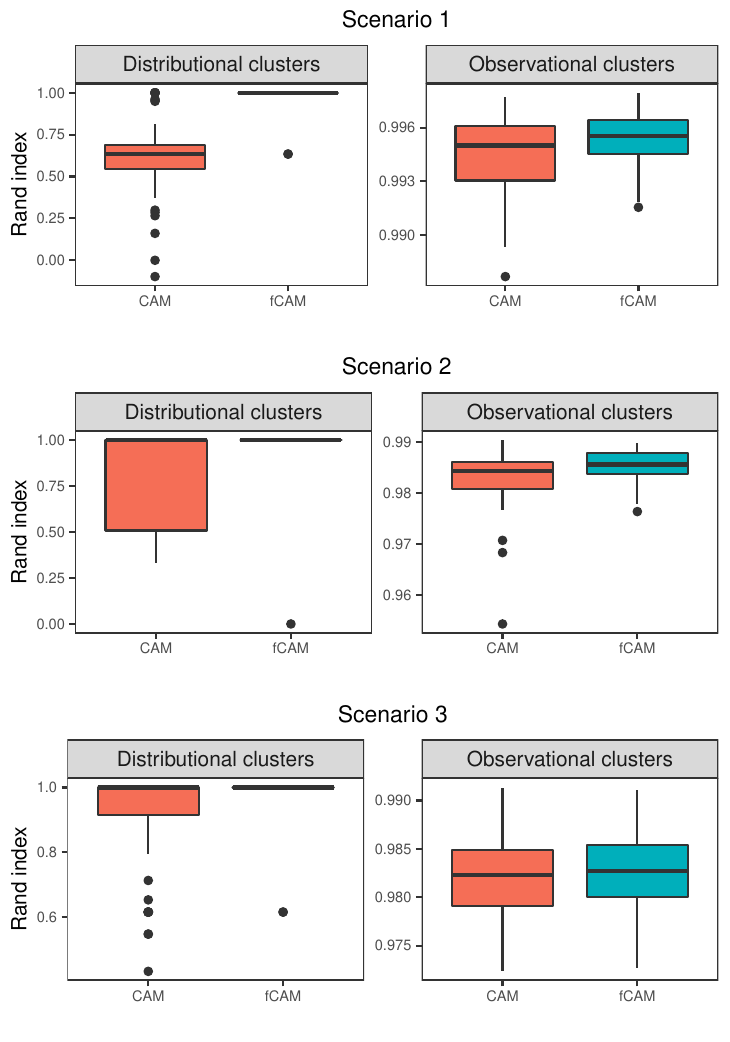}}
	\caption{Distribution of the adjusted Rand index on the distributional and observational clusters, computed on the 50 simulations for the three scenarios of the simulated data. This figure appears in color in the electronic version of this article, and any mention of color refers to that version.}
	\label{fig:rand_index}
\end{figure}

However, the proposed fCAM not only enables the detection of spikes but also allows us to conduct inference on the clustering structure. Therefore, we report on its ability to identify the clustering structure. Figure~\ref{fig:rand_index} reports the adjusted Rand index~\citep{rand1971,hubert1985} computed on both the observational and the distributional clusters for $h_{A1} = h_{A2} = 8$ (results for other settings are similar). Values of the adjusted Rand index close to 1 denote that the identified structure resemble the true clustering. While for the observational clusters the results are broadly comparable, for the distributional clusters, the performance of the proposed fCAM is uniformly superior. In addition, the variability of the results generally appears to be drastically smaller for the fCAM, thus providing evidence of greater efficiency. This is consistent to the results of \citet{fruhwirthschnatter2020} where the generalized mixture of finite mixtures is compared to a standard Dirichlet process mixture model. 

From a computational point of view, the proposed algorithm is clearly more demanding than the optimization methods of~\citet{jewell2019} and~\citet{friedrich2017}.  However, the computing time is comparable to the slice sampler adopted for the CAM, and in general a full run requires just few minutes on a Linux machine with an i7-7700HQ 3.8 GHz Intel processor, 8 GB RAM, running R 4.1.0. For example, for a calcium trace of length 50,000, the computing time of the proposed method is around 2 minutes. Indeed, our experience suggests that the main factor affecting the computing time is the length of the series.  In general, in the analysis of spike activity, we expect the number of clusters  to be small and -- in particular -- much smaller than the number of observations.

\section{Allen Brain Observatory data analysis}
\label{s:dataanalysis}

\begin{figure}
	\centerline{\includegraphics[width = \linewidth]{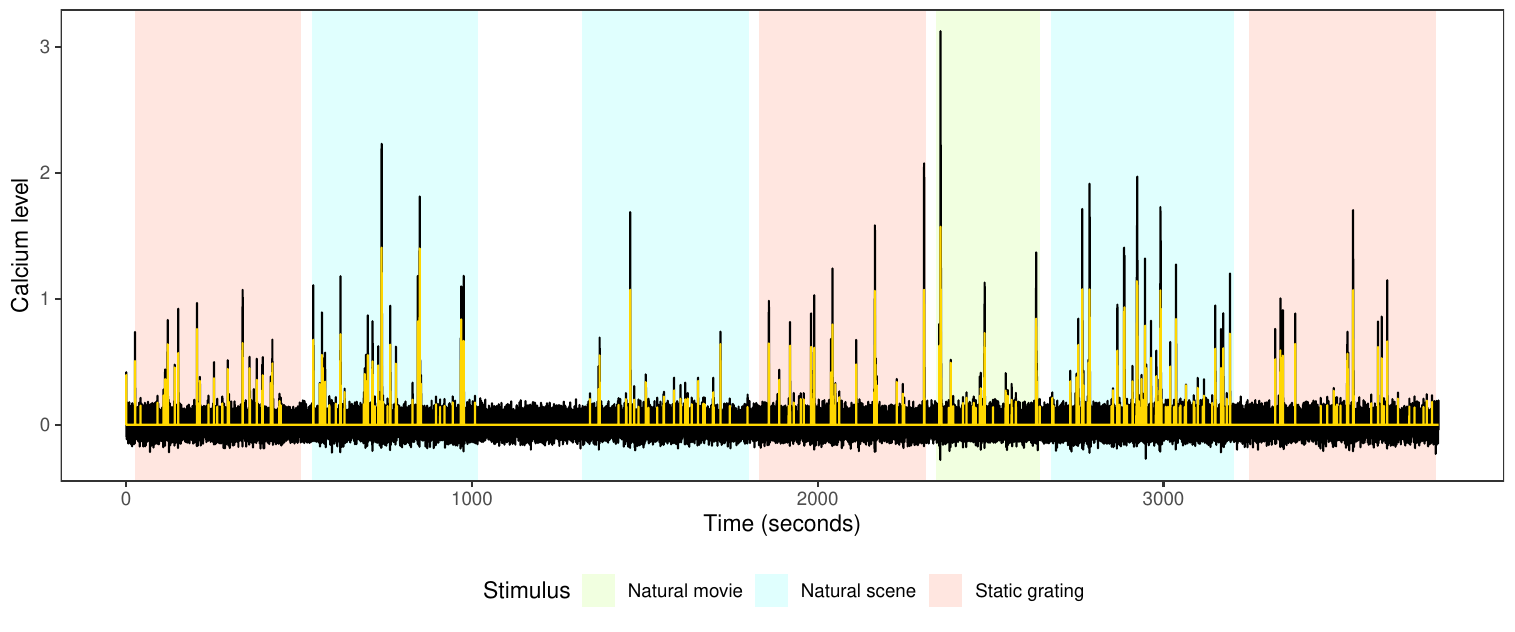}}
	\caption{Observed fluorescence trace $y_t$ from the Allen Brain Observatory data (dark line), and visual stimulus to which the mouse is exposed (shaded areas). The yellow line represents the estimated neuronal activity.}
	\label{fig:y}
\end{figure}

We now revert to the analysis of the data from the Allen Brain Observatory~\citep{allen}. The data comprise the $dF/F$-transformed fluorescence trace for a cell during session-B of the experiment. This session comprises three types of visual stimuli (static gratings, natural scene and natural movie) in addition to some period of spontaneous activity (absence of visual stimuli). Since the data are recorded at a frequency of $30$ Hz, the resulting series consists of 113{,}865 time points for a total of 63.2 minutes. 
We focus our analysis on a neuron located in the primary visual area, at an imaging depth equal to 350 microns (cell id 508596945). Additional analyses for other neurons are reported in the Supporting Information.

The observed fluorescence trace is shown with a continuous black line in Figure~\ref{fig:y}. Different shaded backgrounds indicate the types of visual stimuli.
Using the notation introduced in the previous Sections, $J=4$ with $j=1, 2, 3$ corresponding to static grating, natural scene, and natural movie, respectively and $j=4$ indicating no stimulus presence.

We ran the MCMC algorithm of Section \ref{s:posterior_inference} using the same prior specification of Section \ref{s:model} for 15{,}000 iterations discarding the first 7{,}000 iterations as burn-in and keeping one iteration every four to improve mixing.  Visual inspection of the traceplots and Geweke diagnostics showed no issues with convergence. 
The superimposed light line in Figure~\ref{fig:y} represents the estimated neuronal activity in terms of the inferred amplitude $A_t$, i.e. removing the measurement errors and the result of the accumulation of calcium from the previous spikes. Here and henceforth, we identified the presence of a spike if the  posterior probability of a spike at time $t$, say $PPS_t$, estimated by the proportion of non-zero $A_t$'s over all MCMC iterations, was greater than  $\kappa=75.5\%$. This threshold allows us to control the (estimated) Bayesian false discovery rate at the pre-set value 0.05, that is $\kappa$ solves the equation $\operatorname{FDR}\left(\kappa\right)=\frac{\sum_{t=1}^{T}\left(1-\mathrm{PPS}_{t}\right) I_{\left(\mathrm{PPS}_{t}>\kappa\right)}}{ \left.\sum_{t=1}^{T} I_{\left(\mathrm{PPS}_{t}>\kappa\right.}\right)}=0.05$. For more details, we refer to \citet{newton2004} and \citet{Muller07}. See also \citet{SunReich2015} for a discussion with dependent hypotheses.
As already mentioned in the Introduction, in calcium imaging it is of interest studying the distribution of the spikes in response to each experimental stimulus, and identifying similarities and differences in these distributions across stimuli. 

We start by investigating the presence of similarities in the neuronal response to different types of visual stimuli. This corresponds to analyzing the clustering of the spike distributions induced by the proposed fCAM. The model clusters together the groups corresponding to the natural scene and natural movie stimuli with high posterior probability, while the static grating stimulus and the absence of stimuli are assigned to two separate distributional clusters. In other terms, the neuron appears to show similar neuronal responses in the natural scene and natural movie stimuli whereas the responses appear distinctly different under the other two conditions.

\begin{figure}
	\centerline{\includegraphics[width = .6\linewidth]{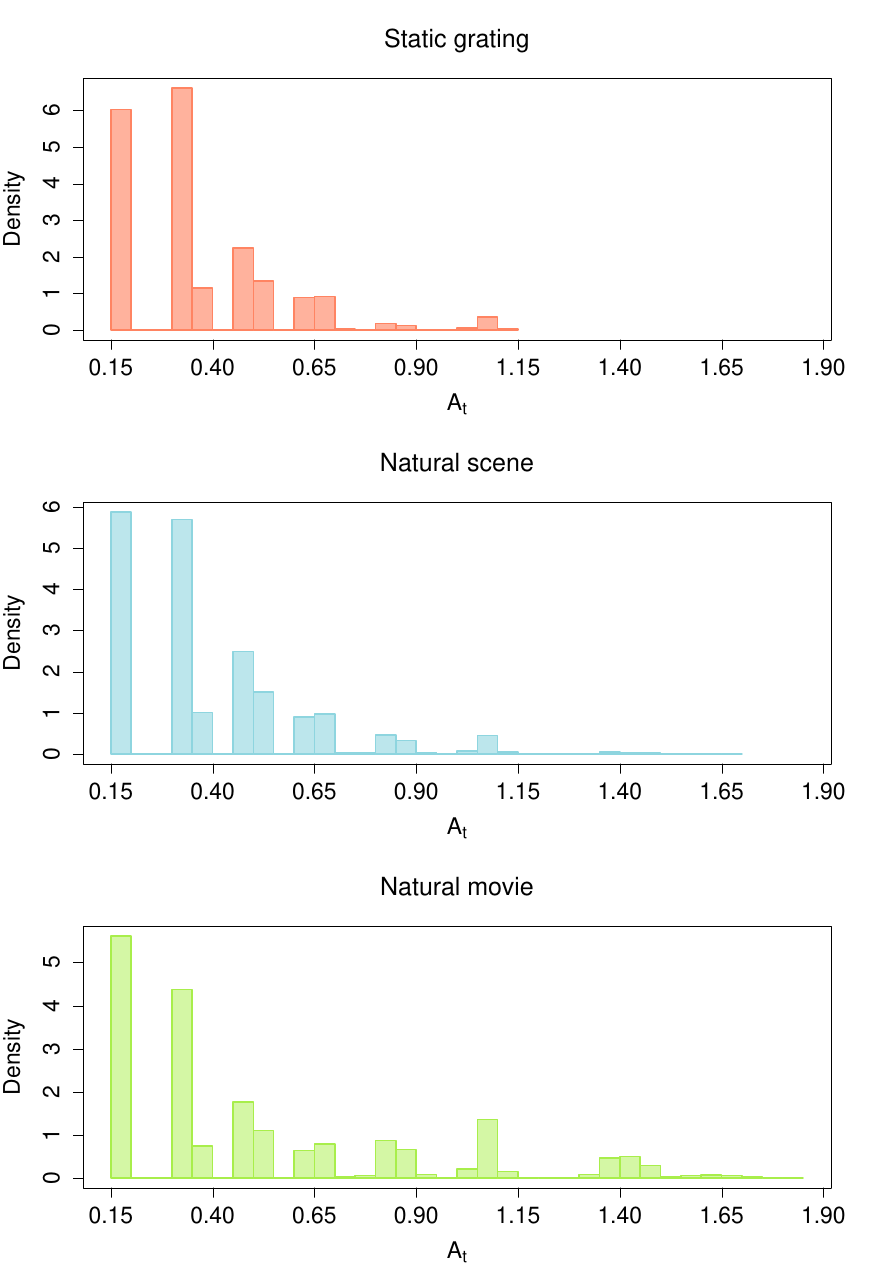}}
	\caption{Empirical distribution of the posterior means of the observational cluster parameters $A_t$ for the three experimental conditions of the Allen Brain Observatory data. This figure appears in color in the electronic version of this article, and any mention of color refers to that version.}
	\label{fig:A_distr}
\end{figure}

To understand whether and how the neuronal response depends on the type of stimulus, we estimated the spike amplitude distribution for each of the four types of stimuli. Figure~\ref{fig:A_distr} shows the histograms of posterior means of the non-zero spike amplitudes for the three types of stimuli. The distribution for the time interval between 1018-1319 sec in Figure \ref{fig:y} (absence of stimuli) is not presented because no activity was detected. Despite the apparent similarities of the distributions in Figure~\ref{fig:A_distr}, the second cluster of spike amplitude distributions (natural scene and natural movie) shows a heavier tail. 
Specifically, the highest observed cluster during the static grating stimulus (top plot) is centered at 1.06, while for the other two stimuli we obtained several higher values, with the largest cluster centered around 1.43. 

\begin{figure}
	\centerline{\includegraphics[width = .6\linewidth]{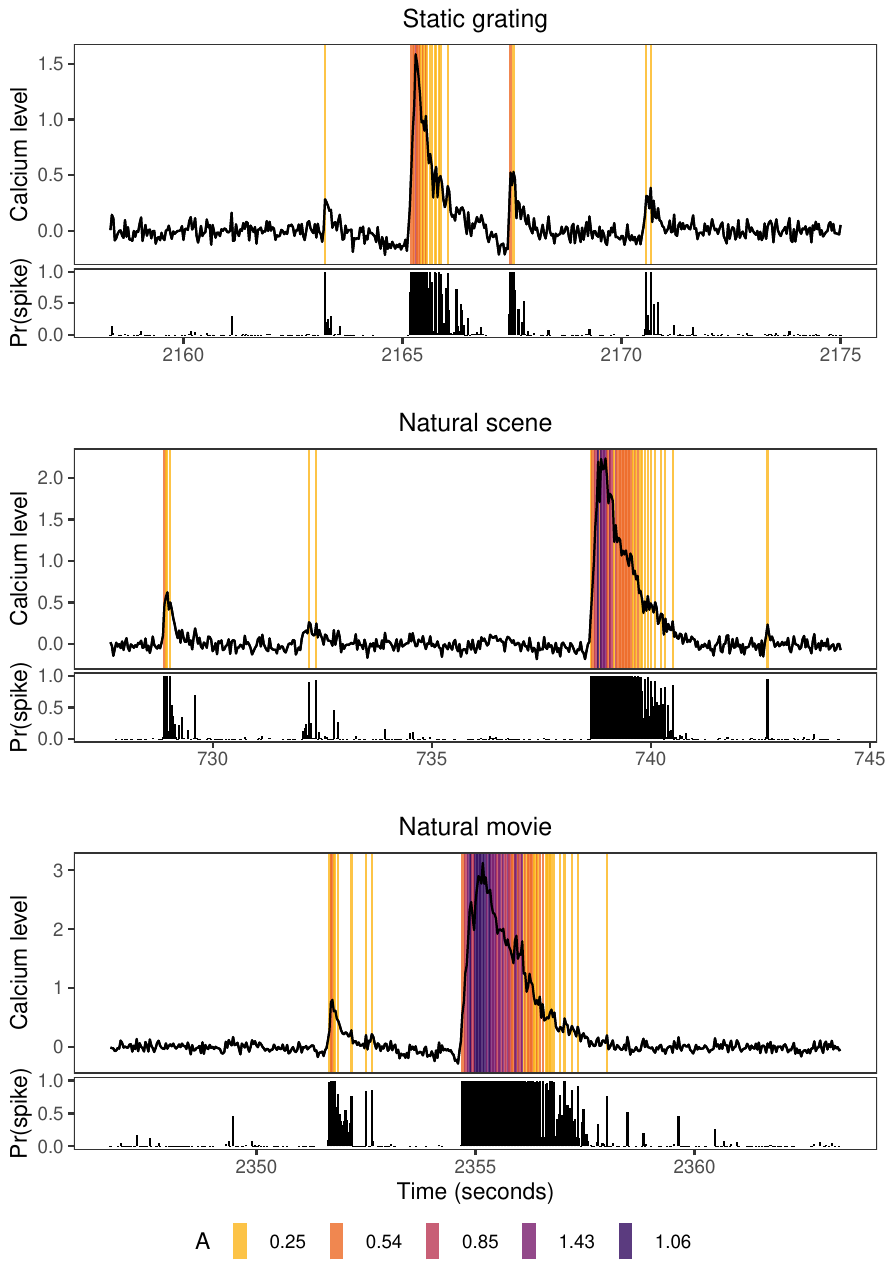}}
	\caption{Short interval of length 500 of the Allen Brain Observatory data in correspondence of a spike, for the three stimuli. The vertical lines indicate the time of a spike and the colors correspond to the observational cluster of its amplitude. The bottom panels show the estimated posterior probability of spike presence, for each time point. }
	\label{fig:spike_color}
\end{figure}

A qualitative representation of how these spike clusters are distributed within the three groups is given in Figure~\ref{fig:spike_color}. The three plots show a short interval of the observed calcium series, chosen in correspondence of one of the highest observed spikes. 
Each plot also shows a series of colored vertical lines: the lines are placed at the estimated spike times, and the colors correspond to the estimated spike amplitudes. 
The represented partition is the posterior point estimate obtained by minimizing the variation of information loss, as proposed in~\citet{wade2018}. Conditionally on the obtained partition, for each cluster a representative value for the cluster parameter is obtained as follows: 
first, for each MCMC iteration, the group-specific average of $A_t$ is computed keeping the partition fixed; then, these values are averaged over all the MCMC iterations.
We notice that for all experiments, high values of the observed calcium level are often produced as the result of several consecutive spikes, since, individually, the spikes are characterized by a relatively low amplitude, and the observed calcium level is cumulated due to its autoregressive behavior.  The autoregressive parameter $\gamma$ has a posterior mean equal to $0.493$ with a 95\% credible interval of $(0.481, 0.505)$.
This result corresponds to the understanding that the observed calcium response may be generated by high-frequency firing neurons: due to the low-sampling rate, the non-linear calcium signal essentially captures a super-imposition of multiple spikes \citep{Hoang2020}.

As a matter of fact, another useful quantity we can compute to compare the neuronal activity between stimuli is the firing rate, which provides a measure of how often the neuron has activated during a specific visual stimulus.
The rate computes the number of detected spikes per second, to take into account the different duration of the experiments.  For the static grating stimulus the posterior mean rate (and related 95\% credible interval) is $0.223$ $(0.216, 0.229)$, while for the natural scene and natural movie stimuli they are $0.419$ $(0.410, 0.428)$ and $0.511$ $(0.495, 0.531)$, respectively. These results highlight the role of spike-frequency adaptation, whereby some neurons show an increased activity when exposed to more complex stimuli, thus exhibiting higher firing rates and larger calcium concentration measurements \citep{Peron2009}.

\section{Discussion}

\label{s:discuss}

Calcium imaging has become widely applied to record the neuronal activity in awake, freely moving and behaving animals. However, reliable spike detection and spike time estimation remain challenging, due to the non-linearity and low signal-to-noise ratio of the calcium response. We have proposed a single-stage nested Bayesian finite mixture model that allows estimating the spike activity and also characterizing how its distribution varies across stimuli. 
The method shows good performances in a simulation study and captures characteristic features of neuronal activity in an application to publicly available data from the Allen Brain Observatory. 

Our approach exploited the knowledge of the stimulus types to model the partial exchangeability  of the data within a Bayesian nonparametric framework. As a referee pointed out,  continuous covariates are often available in the neurosciences, e.g. the position of an alive animal in a two dimensional environment. Our model cannot be directly applied with this information. The inclusion of available continuous and time-varying covariates is the subject of ongoing investigation.

As neuroscience progresses, it is  becoming more apparent that history effects can impact also seemingly low-dimensional experiments as the one considered in the Allen Brain Observatory data. Our approach does not explicitly model spike history dependence. However, it may be possible to explore the effect of history by considering appropriately defined time-windows (e.g., by trials) and fix $J$ equal to the total number of segments,  regardless of our knowledge that some conditions are repeated. If history effects are present, one should expect that even segments corresponding to the same conditions may be assigned with high probability to different clusters as time progresses. However, a proper modeling approach should incorporate further prior constraints (e.g., constraints that take into account the temporal sequence of the segments) to ensure  interpretability of the inference.

In line with the current literature, our approach is limited to the analysis of the calcium responses observed from a single neuron. Methods to infer neuronal connectivity from calcium imaging data over multiple regions of the brain remain sparse, often limited to point estimates \citep{mishchenko2011} or the analysis of \textit{in-vitro} data \citep{rigat2006}. 
Inferences from our work  could possibly be used to identify patterns across multiple neurons. For example, it is reasonable to assume that neurons exhibiting similar activity patterns may be grouped into homogeneous (spatial) clusters. Therefore, a second stage of the analysis may explicitly cluster across neurons the inferred spikes and the posterior means of the amplitudes within successive time-intervals of calcium traces. A bi-clustering approach \citep[see, e.g.,][]{Turner2005, Chekouo2015} could also be employed to  describe the evolution of the neuronal patterns over time and conditions, as well as over space. Alternatively, one could apply the zero-inflated gamma model recently proposed by \citet{wei2019} to study the densities of the deconvolved activity estimates and similarly heuristically compare such densities across neurons. 
Possible extensions of the framework presented here may focus on encoding the dynamic clustering of temporally correlated groups of neurons within the fCAM prior, as a function of external stimuli or the movement of the animal through an environment. 
In addition to the low signal-to-noise ratio, issues related to the dimension of the data and the accurate identification of the locations of neurons further compound the statistical and computational challenges \citep{petersen2018}.

\bibliographystyle{agsm} \bibliography{biblio.bib}

\begin{thebibliography}{}

\bibitem[\protect\citeauthoryear{{Allen Brain Observatory}}{{Allen Brain
  Observatory}}{2017}]{allen_stimulus}
{Allen Brain Observatory} (2017).
\newblock Technical whitepaper: stimulus set and response analyses.
\newblock help.brain-map.org/display/observatory/Data+-+Visual+Coding.

\bibitem[\protect\citeauthoryear{{Allen Institute MindScope Program}}{{Allen
  Institute MindScope Program}}{2016}]{allen}
{Allen Institute MindScope Program} (2016).
\newblock {Allen Brain Observatory} -- 2-photon visual coding [dataset].
\newblock brain-map.org/explore/circuits.

\bibitem[\protect\citeauthoryear{Argiento and {De Iorio}}{Argiento and {De
  Iorio}}{2019}]{argiento2019}
Argiento, R. and {De Iorio}, M. (2019).
\newblock Is infinity that far? {A Bayesian} nonparametric perspective of
  finite mixture models.
\newblock arXiv:1904.09733.

\bibitem[\protect\citeauthoryear{Brenner, Agam, Bialek, and {de Ruyter van
  Steveninck}}{Brenner et~al.}{2002}]{Brenner2002PhysRevE}
Brenner, N., Agam, O., Bialek, W., and {de Ruyter van Steveninck}, R. (2002).
\newblock Statistical properties of spike trains: universal and
  stimulus-dependent aspects.
\newblock {\em Physical review. E, Statistical, nonlinear, and soft matter
  physics} {\bf 66,} 031907.

\bibitem[\protect\citeauthoryear{Camerlenghi, Dunson, Lijoi, Pr{\"u}nster, and
  Rodr{\'\i}guez}{Camerlenghi et~al.}{2019}]{camerlenghi2019}
Camerlenghi, F., Dunson, D.~B., Lijoi, A., Pr{\"u}nster, I., and
  Rodr{\'\i}guez, A. (2019).
\newblock Latent nested nonparametric priors (with discussion).
\newblock {\em Bayesian Analysis} {\bf 14,} 1303--1356.

\bibitem[\protect\citeauthoryear{Canale, Lijoi, Nipoti, and
  Pr{\"u}nster}{Canale et~al.}{2017}]{canale2017}
Canale, A., Lijoi, A., Nipoti, B., and Pr{\"u}nster, I. (2017).
\newblock {On the Pitman--Yor process with spike and slab base measure}.
\newblock {\em Biometrika} {\bf 104,} 681--697.

\bibitem[\protect\citeauthoryear{Canale, Lijoi, Nipoti, and Pr\"unster}{Canale
  et~al.}{2021}]{spikeandslab2}
Canale, A., Lijoi, A., Nipoti, B., and Pr\"unster, I. (2021).
\newblock Inner spike and slab bayesian nonparametric models.
\newblock {\em Econometrics and Statistics} .

\bibitem[\protect\citeauthoryear{Chekouo, Murua, and Raffelsberger}{Chekouo
  et~al.}{2015}]{Chekouo2015}
Chekouo, T., Murua, A., and Raffelsberger, W. (2015).
\newblock {The Gibbs-plaid biclustering model}.
\newblock {\em The Annals of Applied Statistics} {\bf 9,} 1643 -- 1670.

\bibitem[\protect\citeauthoryear{Dana, Sun, Mohar, Hulse, Kerlin, Hasseman,
  Tsegaye, Tsang, Wong, Patel, Macklin, Chen, Konnerth, Jayaraman, Looger,
  Schreiter, Svoboda, and Kim}{Dana et~al.}{2019}]{dana2019}
Dana, H., Sun, Y., Mohar, B., Hulse, B.~K., Kerlin, A.~M., Hasseman, J.~P. et al. (2019).
\newblock High-performance calcium sensors for imaging activity in neuronal
  populations and microcompartments.
\newblock {\em Nature Methods} {\bf 16,} 649 -- 657.

\bibitem[\protect\citeauthoryear{{de Vries}, Lecoq, Buice, Groblewski, Ocker,
  Oliver, Feng, Cain, Ledochowitsch, Millman, Roll, Garrett, Keenan, Kuan,
  Mihalas, Olsen, Thompson, Wakeman, Waters, and Koch}{{de Vries}
  et~al.}{2020}]{vries2020}
{de Vries}, S., Lecoq, J., Buice, M., Groblewski, P., Ocker, G., Oliver, M. et al. (2020).
\newblock A large-scale standardized physiological survey reveals functional
  organization of the mouse visual cortex.
\newblock {\em Nature neuroscience} {\bf 23,} 138--151.

\bibitem[\protect\citeauthoryear{Denti, Camerlenghi, Guindani, and Mira}{Denti
  et~al.}{2021}]{denti2020}
Denti, F., Camerlenghi, F., Guindani, M., and Mira, A. (2021).
\newblock A common atoms model for the {Bayesian} nonparametric analysis of
  nested data.
\newblock {\em Journal of the American Statistical Association} .

\bibitem[\protect\citeauthoryear{Dombeck, Harvey, Tian, Looger, and
  Tank}{Dombeck et~al.}{2010}]{Dombeck2010}
Dombeck, D.~A., Harvey, C.~D., Tian, L., Looger, L.~L., and Tank, D.~W. (2010).
\newblock Functional imaging of hippocampal place cells at cellular resolution
  during virtual navigation.
\newblock {\em Nature Neuroscience} {\bf 13,} 1433 -- 1440.

\bibitem[\protect\citeauthoryear{Friedrich and Paninski}{Friedrich and
  Paninski}{2016}]{friedrich2016}
Friedrich, J. and Paninski, L. (2016).
\newblock Fast active set methods for online spike inference from calcium
  imaging.
\newblock In Lee, D., Sugiyama, M., Luxburg, U., Guyon, I., and Garnett, R.,
  editors, {\em Advances In Neural Information Processing Systems}, pages 1984
  -- 1992.

\bibitem[\protect\citeauthoryear{Friedrich, Zhou, and Paninski}{Friedrich
  et~al.}{2017}]{friedrich2017}
Friedrich, J., Zhou, P., and Paninski, L. (2017).
\newblock Fast online deconvolution of calcium imaging data.
\newblock {\em PLOS Computational Biology} {\bf 13,} 1 -- 26.

\bibitem[\protect\citeauthoryear{Fr\"uhwirth-Schnatter, Malsiner-Walli, and
  Gr\"un}{Fr\"uhwirth-Schnatter et~al.}{2021}]{fruhwirthschnatter2020}
Fr\"uhwirth-Schnatter, S., Malsiner-Walli, G., and Gr\"un, B. (2021).
\newblock Generalized mixtures of finite mixtures and telescoping sampling.
\newblock {\em Bayesian Analysis} {\bf 16,} 1279 -- 1307.

\bibitem[\protect\citeauthoryear{Grienberger and Konnerth}{Grienberger and
  Konnerth}{2012}]{Grienberger2012}
Grienberger, C. and Konnerth, A. (2012).
\newblock Imaging calcium in neurons.
\newblock {\em Neuron} {\bf 73,} 862 -- 885.

\bibitem[\protect\citeauthoryear{Hoang, Sato, Shinomoto, Tsutsumi, Hashizume,
  Ishikawa, Kano, Ikegaya, Kitamura, Kawato, and Toyama}{Hoang
  et~al.}{2020}]{Hoang2020}
Hoang, H., Sato, M., Shinomoto, S., Tsutsumi, S., Hashizume, M., Ishikawa, T. et al. (2020).
\newblock Improved hyperacuity estimation of spike timing from calcium imaging.
\newblock {\em Scientific Reports} {\bf 10,} 17844.

\bibitem[\protect\citeauthoryear{Hubert and Arabie}{Hubert and
  Arabie}{1985}]{hubert1985}
Hubert, L. and Arabie, P. (1985).
\newblock Comparing partitions.
\newblock {\em Journal of Classification} {\bf 2,} 193 -- 218.

\bibitem[\protect\citeauthoryear{Jewell and Witten}{Jewell and
  Witten}{2018}]{jewell2018}
Jewell, S. and Witten, D. (2018).
\newblock Exact spike train inference via {L0} optimization.
\newblock {\em The Annals of Applied Statistics} {\bf 12,} 2457 -- 2482.

\bibitem[\protect\citeauthoryear{Jewell, Hocking, Fearnhead, and Witten}{Jewell
  et~al.}{2019}]{jewell2019}
Jewell, S.~W., Hocking, T.~D., Fearnhead, P., and Witten, D.~M. (2019).
\newblock Fast nonconvex deconvolution of calcium imaging data.
\newblock {\em Biostatistics} {\bf 21,} 709--726.

\bibitem[\protect\citeauthoryear{Johnson and Rossell}{Johnson and
  Rossell}{2010}]{johnson2010}
Johnson, V.~E. and Rossell, D. (2010).
\newblock On the use of non-local prior densities in {Bayesian} hypothesis
  tests.
\newblock {\em Journal of the Royal Statistical Society: Series B (Statistical
  Methodology)} {\bf 72,} 143--170.

\bibitem[\protect\citeauthoryear{Li, Chen, Guo, Gerfen, and Svoboda}{Li
  et~al.}{2015}]{li2015motor}
Li, N., Chen, T.-W., Guo, Z.~V., Gerfen, C.~R., and Svoboda, K. (2015).
\newblock A motor cortex circuit for motor planning and movement.
\newblock {\em Nature} {\bf 519,} 51--56.

\bibitem[\protect\citeauthoryear{Maruyama, Maeda, Moroda, Kato, Inoue,
  Miyakawa, and Aonishi}{Maruyama et~al.}{2014}]{Maruyama2014}
Maruyama, R., Maeda, K., Moroda, H., Kato, I., Inoue, M., Miyakawa, H. et al. (2014).
\newblock Detecting cells using non-negative matrix factorization on calcium
  imaging data.
\newblock {\em Neural Networks} {\bf 55,} 11--19.

\bibitem[\protect\citeauthoryear{Miller and Harrison}{Miller and
  Harrison}{2018}]{miller2018}
Miller, J.~W. and Harrison, M.~T. (2018).
\newblock Mixture models with a prior on the number of components.
\newblock {\em Journal of the American Statistical Association} {\bf 113,}
  340--356.

\bibitem[\protect\citeauthoryear{Mishchenko, Vogelstein, and
  Paninski}{Mishchenko et~al.}{2011}]{mishchenko2011}
Mishchenko, Y., Vogelstein, J.~T., and Paninski, L. (2011).
\newblock A {Bayesian} approach for inferring neuronal connectivity from
  calcium fluorescent imaging data.
\newblock {\em Annals of Applied Statistics} {\bf 5,} 1229--1261.

\bibitem[\protect\citeauthoryear{Mitchell and Beauchamp}{Mitchell and
  Beauchamp}{1988}]{mitchell(88)}
Mitchell, T.~J. and Beauchamp, J.~J. (1988).
\newblock Bayesian variable selection in linear regression.
\newblock {\em Journal of the American Statistical Association} {\bf 83,}
  1023--1036.

\bibitem[\protect\citeauthoryear{Mukamel, Nimmerjahn, and Schnitzer}{Mukamel
  et~al.}{2009}]{Mukamel2009}
Mukamel, E.~A., Nimmerjahn, A., and Schnitzer, M.~J. (2009).
\newblock Automated analysis of cellular signals from large-scale calcium
  imaging data.
\newblock {\em Neuron} {\bf 63,} 747--760.

\bibitem[\protect\citeauthoryear{M\"uller, Parmigiani, and Rice}{M\"uller
  et~al.}{2007}]{Muller07}
M\"uller, P., Parmigiani, G., and Rice, K. (2007).
\newblock {FDR} and {B}ayesian multiple comparisons rules.
\newblock In Bernardo, J., Bayarri, M., Berger, J., Dawid, A., Heckerman, D.,
  Smith, A., and West, M., editors, {\em Bayesian Statistics 8}. Oxford, UK:
  Oxford University Press.

\bibitem[\protect\citeauthoryear{Nakajima and Schmitt}{Nakajima and
  Schmitt}{2020}]{Nakajima2020}
Nakajima, M. and Schmitt, L.~I. (2020).
\newblock Understanding the circuit basis of cognitive functions using mouse
  models.
\newblock {\em Neuroscience Research} {\bf 152,} 44 -- 58.

\bibitem[\protect\citeauthoryear{Newton, Noueiry, Sarkar, and Ahlquist}{Newton
  et~al.}{2004}]{newton2004}
Newton, M.~A., Noueiry, A., Sarkar, D., and Ahlquist, P. (2004).
\newblock Detecting differential gene expression with a semiparametric
  hierarchical mixture method.
\newblock {\em Biostatistics} {\bf 5,}.

\bibitem[\protect\citeauthoryear{Peron and Gabbiani}{Peron and
  Gabbiani}{2009}]{Peron2009}
Peron, S.~P. and Gabbiani, F. (2009).
\newblock Role of spike-frequency adaptation in shaping neuronal response to
  dynamic stimuli.
\newblock {\em Biological cybernetics} {\bf 100,} 505--520.

\bibitem[\protect\citeauthoryear{Petersen, Simon, and Witten}{Petersen
  et~al.}{2018}]{petersen2018}
Petersen, A., Simon, N., and Witten, D. (2018).
\newblock Scalpel: Extracting neurons from calcium imaging data.
\newblock {\em Annals of Applied Statistics} {\bf 12,} 2430--2456.

\bibitem[\protect\citeauthoryear{Pnevmatikakis, Merel, Pakman, and
  Paninski}{Pnevmatikakis et~al.}{2013}]{pnevmatikakis2013}
Pnevmatikakis, E., Merel, J., Pakman, A., and Paninski, L. (2013).
\newblock Bayesian spike inference from calcium imaging data.
\newblock In {\em In Signals, Systems and Computers}, pages 349 -- 353.

\bibitem[\protect\citeauthoryear{Prado and West}{Prado and
  West}{2010}]{prado2010}
Prado, R. and West, M. (2010).
\newblock {\em Time Series: Modeling, Computation, and Inference}.
\newblock Chapman and Hall, 1st edition.

\bibitem[\protect\citeauthoryear{Rand}{Rand}{1971}]{rand1971}
Rand, W.~M. (1971).
\newblock Objective criteria for the evaluation of clustering methods.
\newblock {\em Journal of the American Statistical Association} {\bf 66,}
  846--850.

\bibitem[\protect\citeauthoryear{Rigat, de~Gunst, and van Pelt}{Rigat
  et~al.}{2006}]{rigat2006}
Rigat, F., de~Gunst, M., and van Pelt, J. (2006).
\newblock Bayesian modelling and analysis of spatio-temporal neuronal networks.
\newblock {\em Bayesian Analysis} {\bf 1,} 733--764.

\bibitem[\protect\citeauthoryear{Rodr\'iguez, Dunson, and Gelfand}{Rodr\'iguez
  et~al.}{2008}]{Rodriguez2008}
Rodr\'iguez, A., Dunson, D.~B., and Gelfand, A.~E. (2008).
\newblock The nested {Dirichlet} process.
\newblock {\em Journal of the American Statistical Association} {\bf 103,}
  1131--1154.

\bibitem[\protect\citeauthoryear{Rose, Goltstein, Portugues, and
  Griesbeck}{Rose et~al.}{2014}]{Rose2014}
Rose, T., Goltstein, P.~M., Portugues, R., and Griesbeck, O. (2014).
\newblock Putting a finishing touch on gecis.
\newblock {\em Frontiers in Molecular Neuroscience} {\bf 7,} 88.

\bibitem[\protect\citeauthoryear{Shen, Lur, Xu, and Yu}{Shen
  et~al.}{2021}]{shen2021}
Shen, T., Lur, G., Xu, X., and Yu, Z. (2021).
\newblock To deconvolve, or not to deconvolve: Inferences of neuronal
  activities using calcium imaging data.
\newblock arXiv: 2103.02163.

\bibitem[\protect\citeauthoryear{Shibue and Komaki}{Shibue and
  Komaki}{2020}]{Shibue2020}
Shibue, R. and Komaki, F. (2020).
\newblock Deconvolution of calcium imaging data using marked point processes.
\newblock {\em PLOS Computational Biology} {\bf 16,} 1--25.

\bibitem[\protect\citeauthoryear{Sun, Reich, Tony~Cai, Guindani, and
  Schwartzman}{Sun et~al.}{2015}]{SunReich2015}
Sun, W., Reich, B.~J., Tony~Cai, T., Guindani, M., and Schwartzman, A. (2015).
\newblock False discovery control in large-scale spatial multiple testing.
\newblock {\em Journal of the Royal Statistical Society: Series B (Statistical
  Methodology)} {\bf 77,} 59--83.

\bibitem[\protect\citeauthoryear{Turner, Bailey, and Krzanowski}{Turner
  et~al.}{2005}]{Turner2005}
Turner, H., Bailey, T., and Krzanowski, W. (2005).
\newblock Improved biclustering of microarray data demonstrated through
  systematic performance tests.
\newblock {\em Computational Statistics \& Data Analysis} {\bf 48,} 235--254.

\bibitem[\protect\citeauthoryear{Vogelstein, Packer, Machado, Sippy, Babadi,
  Yuste, and Paninski}{Vogelstein et~al.}{2010}]{vogelstein2010}
Vogelstein, J.~T., Packer, A.~M., Machado, T.~A., Sippy, T., Babadi, B., Yuste,
  R. et al. (2010).
\newblock Fast nonnegative deconvolution for spike train inference from
  population calcium imaging.
\newblock {\em Journal of Neurophysiology} {\bf 104,} 3691--3704.

\bibitem[\protect\citeauthoryear{Vogelstein, Watson, Packer, Yuste, Jedynak,
  and Paninski}{Vogelstein et~al.}{2009}]{vogelstein2009}
Vogelstein, J.~T., Watson, B.~O., Packer, A.~M., Yuste, R., Jedynak, B., and
  Paninski, L. (2009).
\newblock {Spike inference from calcium imaging using sequential Monte Carlo
  methods}.
\newblock {\em Biophysical Journal} {\bf 97,} 636 -- 655.

\bibitem[\protect\citeauthoryear{Wade and Ghahramani}{Wade and
  Ghahramani}{2018}]{wade2018}
Wade, S. and Ghahramani, Z. (2018).
\newblock Bayesian cluster analysis: point estimation and credible balls (with
  discussion).
\newblock {\em Bayesian Analysis} {\bf 13,} 559--626.

\bibitem[\protect\citeauthoryear{Wei, Zhou, Grosmark, Ajabi, Sparks, Zhou,
  Brandon, Losonczy, and Paninski}{Wei et~al.}{2019}]{wei2019}
Wei, X.-X., Zhou, D., Grosmark, A., Ajabi, Z., Sparks, F., Zhou, P. et al. (2019).
\newblock A zero-inflated gamma model for post-deconvolved calcium imaging
  traces.
\newblock {\em bioRxiv: 637652} .

\end{thebibliography}


\begin{thebibliography}{xx}

\harvarditem{{Allen Brain Observatory}}{2017}{allen_stimulus}
{Allen Brain Observatory}  \harvardyearleft 2017\harvardyearright , `Technical
  whitepaper: stimulus set and response analyses',
  help.brain-map.org/display/observatory/Data+-+Visual+Coding.

\harvarditem{{Allen Institute MindScope Program}}{2016}{allen}
{Allen Institute MindScope Program}  \harvardyearleft 2016\harvardyearright ,
  `{Allen Brain Observatory} -- 2-photon visual coding [dataset]',
  brain-map.org/explore/circuits.

\harvarditem{Argiento \harvardand\ {De Iorio}}{2019}{argiento2019}
Argiento, R. \harvardand\ {De Iorio}, M.  \harvardyearleft
  2019\harvardyearright , `Is infinity that far? a {Bayesian} nonparametric
  perspective of finite mixture models', arXiv:1904.09733.

\harvarditem[Brenner et~al.]{Brenner, Agam, Bialek \harvardand\ {de Ruyter van
  Steveninck}}{2002}{Brenner2002PhysRevE}
Brenner, N., Agam, O., Bialek, W. \harvardand\ {de Ruyter van Steveninck}, R.
  \harvardyearleft 2002\harvardyearright , `Statistical properties of spike
  trains: universal and stimulus-dependent aspects', {\em Physical review. E,
  Statistical, nonlinear, and soft matter physics} {\bf 66},~031907.

\harvarditem[Camerlenghi et~al.]{Camerlenghi, Dunson, Lijoi, Pr{\"u}nster
  \harvardand\ Rodr{\'\i}guez}{2019}{camerlenghi2019}
Camerlenghi, F., Dunson, D.~B., Lijoi, A., Pr{\"u}nster, I. \harvardand\
  Rodr{\'\i}guez, A.  \harvardyearleft 2019\harvardyearright , `Latent nested
  nonparametric priors (with discussion)', {\em Bayesian Analysis} {\bf
  14}(4),~1303--1356.
\newline\harvardurl{https://doi.org/10.1214/19-BA1169}

\harvarditem[Canale et~al.]{Canale, Lijoi, Nipoti \harvardand\
  Pr{\"u}nster}{2017}{canale2017}
Canale, A., Lijoi, A., Nipoti, B. \harvardand\ Pr{\"u}nster, I.
  \harvardyearleft 2017\harvardyearright , `{On the Pitman--Yor process with
  spike and slab base measure}', {\em Biometrika} {\bf 104}(3),~681--697.

\harvarditem[Canale et~al.]{Canale, Lijoi, Nipoti \harvardand\
  Pr\"unster}{2021}{spikeandslab2}
Canale, A., Lijoi, A., Nipoti, B. \harvardand\ Pr\"unster, I.  \harvardyearleft
  2021\harvardyearright , `Inner spike and slab bayesian nonparametric models',
  {\em Econometrics and Statistics} .
\newline\harvardurl{https://www.sciencedirect.com/science/article/pii/S245230622100143X}

\harvarditem[Chekouo et~al.]{Chekouo, Murua \harvardand\
  Raffelsberger}{2015}{Chekouo2015}
Chekouo, T., Murua, A. \harvardand\ Raffelsberger, W.  \harvardyearleft
  2015\harvardyearright , `{The Gibbs-plaid biclustering model}', {\em The
  Annals of Applied Statistics} {\bf 9}(3),~1643 -- 1670.

\harvarditem[Dana et~al.]{Dana, Sun, Mohar, Hulse, Kerlin, Hasseman, Tsegaye,
  Tsang, Wong, Patel, Macklin, Chen, Konnerth, Jayaraman, Looger, Schreiter,
  Svoboda \harvardand\ Kim}{2019}{dana2019}
Dana, H., Sun, Y., Mohar, B., Hulse, B.~K., Kerlin, A.~M., Hasseman, J.~P.,
  Tsegaye, G., Tsang, A., Wong, A., Patel, R., Macklin, J.~J., Chen, Y.,
  Konnerth, A., Jayaraman, V., Looger, L.~L., Schreiter, E.~R., Svoboda, K.
  \harvardand\ Kim, D.~S.  \harvardyearleft 2019\harvardyearright ,
  `High-performance calcium sensors for imaging activity in neuronal
  populations and microcompartments', {\em Nature Methods} {\bf 16},~649 --
  657.

\harvarditem[{de Vries} et~al.]{{de Vries}, Lecoq, Buice, Groblewski, Ocker,
  Oliver, Feng, Cain, Ledochowitsch, Millman, Roll, Garrett, Keenan, Kuan,
  Mihalas, Olsen, Thompson, Wakeman, Waters \harvardand\ Koch}{2020}{vries2020}
{de Vries}, S., Lecoq, J., Buice, M., Groblewski, P., Ocker, G., Oliver, M.,
  Feng, D., Cain, N., Ledochowitsch, P., Millman, D., Roll, K., Garrett, M.,
  Keenan, T., Kuan, C., Mihalas, S., Olsen, S., Thompson, C., Wakeman, W.,
  Waters, J. \harvardand\ Koch, C.  \harvardyearleft 2020\harvardyearright , `A
  large-scale standardized physiological survey reveals functional organization
  of the mouse visual cortex', {\em Nature neuroscience} {\bf 23}(1),~138--151.

\harvarditem[Denti et~al.]{Denti, Camerlenghi, Guindani \harvardand\
  Mira}{2021}{denti2020}
Denti, F., Camerlenghi, F., Guindani, M. \harvardand\ Mira, A.
  \harvardyearleft 2021\harvardyearright , `A common atoms model for the
  {Bayesian} nonparametric analysis of nested data', {\em Journal of the
  American Statistical Association} .

\harvarditem[Dombeck et~al.]{Dombeck, Harvey, Tian, Looger \harvardand\
  Tank}{2010}{Dombeck2010}
Dombeck, D.~A., Harvey, C.~D., Tian, L., Looger, L.~L. \harvardand\ Tank, D.~W.
   \harvardyearleft 2010\harvardyearright , `Functional imaging of hippocampal
  place cells at cellular resolution during virtual navigation', {\em Nature
  Neuroscience} {\bf 13},~1433 -- 1440.

\harvarditem{Friedrich \harvardand\ Paninski}{2016}{friedrich2016}
Friedrich, J. \harvardand\ Paninski, L.  \harvardyearleft 2016\harvardyearright
  , Fast active set methods for online spike inference from calcium imaging,
  {\em in} D.~Lee, M.~Sugiyama, U.~Luxburg, I.~Guyon \harvardand\ R.~Garnett,
  eds, `Advances In Neural Information Processing Systems', pp.~1984 -- 1992.

\harvarditem[Friedrich et~al.]{Friedrich, Zhou \harvardand\
  Paninski}{2017}{friedrich2017}
Friedrich, J., Zhou, P. \harvardand\ Paninski, L.  \harvardyearleft
  2017\harvardyearright , `Fast online deconvolution of calcium imaging data',
  {\em PLOS Computational Biology} {\bf 13}(3),~1 -- 26.

\harvarditem[Fr\"uhwirth-Schnatter et~al.]{Fr\"uhwirth-Schnatter,
  Malsiner-Walli \harvardand\ Gr\"un}{2021}{fruhwirthschnatter2020}
Fr\"uhwirth-Schnatter, S., Malsiner-Walli, G. \harvardand\ Gr\"un, B.
  \harvardyearleft 2021\harvardyearright , `Generalized mixtures of finite
  mixtures and telescoping sampling', {\em Bayesian Analysis} {\bf 16}(4),~1279
  -- 1307.

\harvarditem{Grienberger \harvardand\ Konnerth}{2012}{Grienberger2012}
Grienberger, C. \harvardand\ Konnerth, A.  \harvardyearleft
  2012\harvardyearright , `Imaging calcium in neurons', {\em Neuron} {\bf
  73}(5),~862 -- 885.

\harvarditem[Hoang et~al.]{Hoang, Sato, Shinomoto, Tsutsumi, Hashizume,
  Ishikawa, Kano, Ikegaya, Kitamura, Kawato \harvardand\
  Toyama}{2020}{Hoang2020}
Hoang, H., Sato, M., Shinomoto, S., Tsutsumi, S., Hashizume, M., Ishikawa, T.,
  Kano, M., Ikegaya, Y., Kitamura, K., Kawato, M. \harvardand\ Toyama, K.
  \harvardyearleft 2020\harvardyearright , `Improved hyperacuity estimation of
  spike timing from calcium imaging', {\em Scientific Reports} {\bf
  10}(1),~17844.

\harvarditem{Hubert \harvardand\ Arabie}{1985}{hubert1985}
Hubert, L. \harvardand\ Arabie, P.  \harvardyearleft 1985\harvardyearright ,
  `Comparing partitions', {\em Journal of Classification} {\bf 2}(336),~193 --
  218.

\harvarditem[Jewell et~al.]{Jewell, Hocking, Fearnhead \harvardand\
  Witten}{2019}{jewell2019}
Jewell, S.~W., Hocking, T.~D., Fearnhead, P. \harvardand\ Witten, D.~M.
  \harvardyearleft 2019\harvardyearright , `Fast nonconvex deconvolution of
  calcium imaging data', {\em Biostatistics} {\bf 21}(4),~709--726.

\harvarditem{Jewell \harvardand\ Witten}{2018}{jewell2018}
Jewell, S. \harvardand\ Witten, D.  \harvardyearleft 2018\harvardyearright ,
  `Exact spike train inference via {L0} optimization', {\em The Annals of
  Applied Statistics} {\bf 12}(4),~2457 -- 2482.

\harvarditem{Johnson \harvardand\ Rossell}{2010}{johnson2010}
Johnson, V.~E. \harvardand\ Rossell, D.  \harvardyearleft 2010\harvardyearright
  , `On the use of non-local prior densities in {Bayesian} hypothesis tests',
  {\em Journal of the Royal Statistical Society: Series B (Statistical
  Methodology)} {\bf 72}(2),~143--170.

\harvarditem[Li et~al.]{Li, Chen, Guo, Gerfen \harvardand\
  Svoboda}{2015}{li2015motor}
Li, N., Chen, T.-W., Guo, Z.~V., Gerfen, C.~R. \harvardand\ Svoboda, K.
  \harvardyearleft 2015\harvardyearright , `A motor cortex circuit for motor
  planning and movement', {\em Nature} {\bf 519}(7541),~51--56.

\harvarditem[Maruyama et~al.]{Maruyama, Maeda, Moroda, Kato, Inoue, Miyakawa
  \harvardand\ Aonishi}{2014}{Maruyama2014}
Maruyama, R., Maeda, K., Moroda, H., Kato, I., Inoue, M., Miyakawa, H.
  \harvardand\ Aonishi, T.  \harvardyearleft 2014\harvardyearright , `Detecting
  cells using non-negative matrix factorization on calcium imaging data', {\em
  Neural Networks} {\bf 55},~11--19.

\harvarditem{Miller \harvardand\ Harrison}{2018}{miller2018}
Miller, J.~W. \harvardand\ Harrison, M.~T.  \harvardyearleft
  2018\harvardyearright , `Mixture models with a prior on the number of
  components', {\em Journal of the American Statistical Association} {\bf
  113}(521),~340--356.

\harvarditem[Mishchenko et~al.]{Mishchenko, Vogelstein \harvardand\
  Paninski}{2011}{mishchenko2011}
Mishchenko, Y., Vogelstein, J.~T. \harvardand\ Paninski, L.  \harvardyearleft
  2011\harvardyearright , `A {Bayesian} approach for inferring neuronal
  connectivity from calcium fluorescent imaging data', {\em Annals of Applied
  Statistics} {\bf 5}(2B),~1229--1261.
\newline\harvardurl{https://doi.org/10.1214/09-AOAS303}

\harvarditem{Mitchell \harvardand\ Beauchamp}{1988}{mitchell(88)}
Mitchell, T.~J. \harvardand\ Beauchamp, J.~J.  \harvardyearleft
  1988\harvardyearright , `Bayesian variable selection in linear regression',
  {\em Journal of the American Statistical Association} {\bf
  83}(404),~1023--1036.

\harvarditem[Mukamel et~al.]{Mukamel, Nimmerjahn \harvardand\
  Schnitzer}{2009}{Mukamel2009}
Mukamel, E.~A., Nimmerjahn, A. \harvardand\ Schnitzer, M.~J.  \harvardyearleft
  2009\harvardyearright , `Automated analysis of cellular signals from
  large-scale calcium imaging data', {\em Neuron} {\bf 63}(6),~747--760.

\harvarditem[M\"uller et~al.]{M\"uller, Parmigiani \harvardand\
  Rice}{2007}{Muller07}
M\"uller, P., Parmigiani, G. \harvardand\ Rice, K.  \harvardyearleft
  2007\harvardyearright , {FDR} and {B}ayesian multiple comparisons rules, {\em
  in} J.~Bernardo, M.~Bayarri, J.~Berger, A.~Dawid, D.~Heckerman, A.~Smith
  \harvardand\ M.~West, eds, `Bayesian Statistics 8', Oxford, UK: Oxford
  University Press.

\harvarditem{Nakajima \harvardand\ Schmitt}{2020}{Nakajima2020}
Nakajima, M. \harvardand\ Schmitt, L.~I.  \harvardyearleft
  2020\harvardyearright , `Understanding the circuit basis of cognitive
  functions using mouse models', {\em Neuroscience Research} {\bf 152},~44 --
  58.

\harvarditem[Newton et~al.]{Newton, Noueiry, Sarkar \harvardand\
  Ahlquist}{2004}{newton2004}
Newton, M.~A., Noueiry, A., Sarkar, D. \harvardand\ Ahlquist, P.
  \harvardyearleft 2004\harvardyearright , `Detecting differential gene
  expression with a semiparametric hierarchical mixture method', {\em
  Biostatistics} {\bf 5}(2).

\harvarditem{Peron \harvardand\ Gabbiani}{2009}{Peron2009}
Peron, S.~P. \harvardand\ Gabbiani, F.  \harvardyearleft 2009\harvardyearright
  , `Role of spike-frequency adaptation in shaping neuronal response to dynamic
  stimuli', {\em Biological cybernetics} {\bf 100}(6),~505--520.

\harvarditem[Petersen et~al.]{Petersen, Simon \harvardand\
  Witten}{2018}{petersen2018}
Petersen, A., Simon, N. \harvardand\ Witten, D.  \harvardyearleft
  2018\harvardyearright , `Scalpel: Extracting neurons from calcium imaging
  data', {\em Annals of Applied Statistics} {\bf 12}(4),~2430--2456.
\newline\harvardurl{https://doi.org/10.1214/18-AOAS1159}

\harvarditem[Pnevmatikakis et~al.]{Pnevmatikakis, Merel, Pakman \harvardand\
  Paninski}{2013}{pnevmatikakis2013}
Pnevmatikakis, E., Merel, J., Pakman, A. \harvardand\ Paninski, L.
  \harvardyearleft 2013\harvardyearright , Bayesian spike inference from
  calcium imaging data, {\em in} `In Signals, Systems and Computers', pp.~349
  -- 353.

\harvarditem{Prado \harvardand\ West}{2010}{prado2010}
Prado, R. \harvardand\ West, M.  \harvardyearleft 2010\harvardyearright , {\em
  Time Series: Modeling, Computation, and Inference}, 1st edn, Chapman and
  Hall.

\harvarditem{Rand}{1971}{rand1971}
Rand, W.~M.  \harvardyearleft 1971\harvardyearright , `Objective criteria for
  the evaluation of clustering methods', {\em Journal of the American
  Statistical Association} {\bf 66}(336),~846--850.

\harvarditem[Rigat et~al.]{Rigat, de~Gunst \harvardand\ van
  Pelt}{2006}{rigat2006}
Rigat, F., de~Gunst, M. \harvardand\ van Pelt, J.  \harvardyearleft
  2006\harvardyearright , `Bayesian modelling and analysis of spatio-temporal
  neuronal networks', {\em Bayesian Analysis} {\bf 1}(4),~733--764.
\newline\harvardurl{https://doi.org/10.1214/06-BA124}

\harvarditem[Rodr\'iguez et~al.]{Rodr\'iguez, Dunson \harvardand\
  Gelfand}{2008}{Rodriguez2008}
Rodr\'iguez, A., Dunson, D.~B. \harvardand\ Gelfand, A.~E.  \harvardyearleft
  2008\harvardyearright , `The nested {Dirichlet} process', {\em Journal of the
  American Statistical Association} {\bf 103}(483),~1131--1154.

\harvarditem[Rose et~al.]{Rose, Goltstein, Portugues \harvardand\
  Griesbeck}{2014}{Rose2014}
Rose, T., Goltstein, P.~M., Portugues, R. \harvardand\ Griesbeck, O.
  \harvardyearleft 2014\harvardyearright , `Putting a finishing touch on
  gecis', {\em Frontiers in Molecular Neuroscience} {\bf 7},~88.

\harvarditem[Shen et~al.]{Shen, Lur, Xu \harvardand\ Yu}{2021}{shen2021}
Shen, T., Lur, G., Xu, X. \harvardand\ Yu, Z.  \harvardyearleft
  2021\harvardyearright , `To deconvolve, or not to deconvolve: Inferences of
  neuronal activities using calcium imaging data', arXiv: 2103.02163.

\harvarditem{Shibue \harvardand\ Komaki}{2020}{Shibue2020}
Shibue, R. \harvardand\ Komaki, F.  \harvardyearleft 2020\harvardyearright ,
  `Deconvolution of calcium imaging data using marked point processes', {\em
  PLOS Computational Biology} {\bf 16}(3),~1--25.

\harvarditem[Sun et~al.]{Sun, Reich, Tony~Cai, Guindani \harvardand\
  Schwartzman}{2015}{SunReich2015}
Sun, W., Reich, B.~J., Tony~Cai, T., Guindani, M. \harvardand\ Schwartzman, A.
  \harvardyearleft 2015\harvardyearright , `False discovery control in
  large-scale spatial multiple testing', {\em Journal of the Royal Statistical
  Society: Series B (Statistical Methodology)} {\bf 77}(1),~59--83.

\harvarditem[Turner et~al.]{Turner, Bailey \harvardand\
  Krzanowski}{2005}{Turner2005}
Turner, H., Bailey, T. \harvardand\ Krzanowski, W.  \harvardyearleft
  2005\harvardyearright , `Improved biclustering of microarray data
  demonstrated through systematic performance tests', {\em Computational
  Statistics \& Data Analysis} {\bf 48}(2),~235--254.

\harvarditem[Vogelstein et~al.]{Vogelstein, Packer, Machado, Sippy, Babadi,
  Yuste \harvardand\ Paninski}{2010}{vogelstein2010}
Vogelstein, J.~T., Packer, A.~M., Machado, T.~A., Sippy, T., Babadi, B., Yuste,
  R. \harvardand\ Paninski, L.  \harvardyearleft 2010\harvardyearright , `Fast
  nonnegative deconvolution for spike train inference from population calcium
  imaging', {\em Journal of Neurophysiology} {\bf 104}(6),~3691--3704.

\harvarditem[Vogelstein et~al.]{Vogelstein, Watson, Packer, Yuste, Jedynak
  \harvardand\ Paninski}{2009}{vogelstein2009}
Vogelstein, J.~T., Watson, B.~O., Packer, A.~M., Yuste, R., Jedynak, B.
  \harvardand\ Paninski, L.  \harvardyearleft 2009\harvardyearright , `{Spike
  inference from calcium imaging using sequential Monte Carlo methods}', {\em
  Biophysical Journal} {\bf 97}(2),~636 -- 655.

\harvarditem{Wade \harvardand\ Ghahramani}{2018}{wade2018}
Wade, S. \harvardand\ Ghahramani, Z.  \harvardyearleft 2018\harvardyearright ,
  `Bayesian cluster analysis: point estimation and credible balls (with
  discussion)', {\em Bayesian Analysis} {\bf 13}(2),~559--626.
\newline\harvardurl{https://doi.org/10.1214/17-BA1073}

\harvarditem[Wei et~al.]{Wei, Zhou, Grosmark, Ajabi, Sparks, Zhou, Brandon,
  Losonczy \harvardand\ Paninski}{2019}{wei2019}
Wei, X.-X., Zhou, D., Grosmark, A., Ajabi, Z., Sparks, F., Zhou, P., Brandon,
  M., Losonczy, A. \harvardand\ Paninski, L.  \harvardyearleft
  2019\harvardyearright , `A zero-inflated gamma model for post-deconvolved
  calcium imaging traces', {\em bioRxiv: 637652} .

\end{thebibliography}
 
\appendix
\section{Nested telescoping sampling}
\label{ss:nested_telescoping}
Denote with $\mathcal{C}^D$ the current partition on the distributions and with $\mathcal{C}^O$ the partition on the observations.
\begin{enumerate}
    \item Sample the weights on the distributions: $(\pi_1,\dots,\pi_K)\mid K, \alpha, \mathcal{C}^D \sim \mathrm{Dir}(e_1, \dots, e_K)$; where $e_k = \alpha/K + J_k$, and $J_k$ is the number of groups assigned to the distribution $k$.
    \item Sample the weights on the observations: for all $k\in\{1,\dots,K\}$ sample a vector $\bm{\omega}_k$ from
    $(\omega_{1,k},\dots,\omega_{L,k})\mid L, \beta, \mathcal{C}^O, \mathcal{C}^D \sim \mathrm{Dir}(f_{1,k}, \dots, f_{L,k})$; where $f_{l,k} = \beta/L + N_{l,k}$, and $N_{l,k}$ is the number of observations in the observational cluster $l$ and distributional cluster $k$.
    \item Update the partition on the distributions $\mathcal{C}^D$ by sampling from the posterior distribution of the latent cluster allocation variables $\bm{S}$. For $j = 1,\dots,J$
    $$\mathrm{Pr}(S_j = k\mid \bm{\pi}, K,\bm{A}^*, \bm{y}, \bm{g}) \propto \pi_k \prod_{t:g_t=j} \omega_{M_t,S_j} \: p(y_t\mid A^*_{M_t}),$$
    with $k\in\{1,\dots,K\}$.
    Determine $J_k = \#\{j:S_j = k\}$, for $k=1\dots,K$, and the number of non-empty components $K_+ = \sum_{k=1}^K I\{ J_k > 0\}$. Relabel the components so that the first $K_+$ are non-empty.
    \item Update the partition on the observations $\mathcal{C}^O$ by sampling from the posterior distribution of the latent cluster allocation variables $\bm{M}$. For $t = 1,\dots,T$
    $$\mathrm{Pr}(M_t = l \mid S_{g_t}= k, \bm{S},\bm{\omega}, L,K,\bm{A}^*, \bm{y}, \bm{g}) \propto \omega_{l,k} \: p(y_t\mid A^*_{M_t}),$$
    with $l\in\{1,\dots,L\}$, $k\in\{1,\dots,K\}$.
    Determine $N_l = \#\{t:M_t = l\}$, for $l=1\dots,L$, and the number of non-empty components $L_+ = \sum_{l=1}^L I\{ N_l > 0\}$. Relabel the components so that the first $L_+$ are non-empty. Because all the mixtures share the same atoms, the cluster parameters are sorted regardless of the distributional cluster allocation.
    \item Sample the cluster parameters for the non-empty components: $ p(A^*_l\mid -)\propto p(A^*_l)\: \prod_{t:M_t = l} p(y_t\mid A^*_l) $.
    \item Conditional on $\mathcal{C}^D$, sample the number of components $K$ of the mixture on distributions.
    \item Conditional on $\mathcal{C}^O$, sample the number of components $L$ of the mixtures on observations. If $L > L_+$, sample a new parameter $A^*$ for the empty components from the prior distribution.
    \item Update the hyperparameter $\alpha$ on the Dirichlet distribution on the mixture weights on distributions.
    \item Update the hyperparameter $\beta$ on the Dirichlet distribution on the mixture weights on observations.
\end{enumerate}
The posterior distributions for steps 6-9 are given in~\citet{fruhwirthschnatter2020}.

\end{document}